# New insights on the binary asteroid 121 Hermione[†]

P. Descamps[1], F. Marchis[1,2,11], J. Durech[3], J. Emery[4], A.W. Harris[5], M. Kaasalainen[6], J. Berthier[1], J.-P. Teng-Chuen-Yu[7], A. Peyrot[7], L. Hutton[8], J. Greene[8], J. Pollock[8], M. Assafin[9], R. Vieira-Martins[10], J.I.B. Camargo[9], F. Braga-Ribas[9], F. Vachier[1], D.E. Reichart[12], K.M. Ivarsen[12], J.A. Crain[12], M.C. Nysewander[12], A. P. Lacluyze[12], J.B. Haislip[12], R. Behrend[13], F. Colas[1], J. Lecacheux[14], L. Bernasconi[15], R. Roy[16], P. Baudouin[17], L. Brunetto[18], S. Sposetti[19], F. Manzini[20]

[1] Institut de Mécanique Céleste et de Calcul des Éphémérides, Observatoire de Paris, UMR8028 CNRS, 77 av. Denfert-Rochereau 75014 Paris, France
[2] University of California at Berkeley, Department of Astronomy, 601 Campbell Hall, Berkeley, CA 94720, USA
[3] Astronomical Institute, Charles University in Prague, Faculty of Mathematics and Physics, V Holesovickach 2, 18000 Prague, Czech Republic
[4] University of Tenessee at Knoxville, 306 EPS Building, 1412 Circle Drive, Knoxville, TN 37996, USA
[5] DLR Institute of Planetary Research, Rutherfordstrasse 2, 12489 Berlin, Germany
[6] Department of Mathematics and Statistics, Gustaf Hallstromin katu 2b, P.O.Box 68, FIN-00014 University of Helsinki, Finland
[7] Makes Observatory, 18, rue G. Bizet - Les Makes - 97421 La Rivière, France
[8] Appalachian State University, Department of Physics and Astronomy, 231 CAP Building, Boone, NC 28608, USA
[9] Observatório do Valongo/UFRJ, Ladeira Pedro Antonio 43, 20080-090 Rio de Janeiro, Brazil.
[10] Observatório Nacional, Rua General Jose Cristino 77, 20921-400 Rio de Janeiro, Brazil.
[11] SETI Institute, 515 N. Whisman Road, Mountain View CA 94043, USA
[12] University of North Carolina, Chapel Hill, NC, USA
[13] Geneva Observatory, 1290 Sauverny, Switzerland
[14] Observatoire de Paris, LESIA, 5, Place Jules Janssen, 92195 Meudon cedex, France
[15] Les Engarouines Observatory, 84570 Mallemort-du-Comtat, France
[16] Blauvac Observatory, 84570 St-Estève, France
[17] Harfleur Observatory, 51, Rue Robert Ancel, 76700 Harfleur, France
[18] "Le Florian", Villa 4, 880 chemin de Ribac-Estagnol, 06600 Antibes, France
[19] Gnosca Observatory, 6525 Gnosca, Switzerland
[20] Stazione Astronomica di Sozzago, 28060 Sozzago, Italy

Pages: 39

Tables: 4

Figures: 8

---

[†] Based on observations from program 077.C-0422 collected at the European Southern Observatory, Chile

*Corresponding author:*
**Pascal Descamps**
IMCCE, Paris Observatory
77, avenue Denfert-Rochereau
75014 Paris
France

descamps@imcce.fr
Phone: +33140512268
Fax:     +33146332834

## Abstract

We report on the results of a six-month photometric study of the main-belt binary C-type asteroid 121 Hermione, performed during its 2007 opposition. We took advantage of the rare observational opportunity afforded by one of the annual equinoxes of Hermione occurring close to its opposition in June 2007. The equinox provides an edge-on aspect for an Earth-based observer, which is well suited to a thorough study of Hermione's physical characteristics. The catalog of observations carried out with small telescopes is presented in this work, together with new adaptive optics (AO) imaging obtained between 2005 and 2008 with the Yepun 8-m VLT telescope and the 10-m Keck telescope. The most striking result is confirmation that Hermione is a bifurcated and elongated body, as suggested by Marchis et al., (2005). A new effective diameter of 187 ± 6 km was calculated from the combination of AO, photometric and thermal observations. The new diameter is some 10% smaller than the hitherto accepted radiometric diameter based on IRAS data. The reason for the discrepancy is that IRAS viewed the system almost pole-on. New thermal observations with the Spitzer Space Telescope agree with the diameter derived from AO and lightcurve observations. On the basis of the new AO astrometric observations of the small 32-km diameter satellite we have refined the orbit solution and derived a new value of the bulk density of Hermione of 1.4 +0.5/-0.2 g cm$^{-3}$. We infer a macroscopic porosity of ~33 +5/-20%.

## 1. Introduction

121 Hermione is a large main belt binary asteroid classified as a C-type in the Tholen et al. (1989) taxonomy or Ch-type in the Bus and Binzel (2002) taxonomy. The diameter from IRAS thermal IR measurements was given by Tedesco et al. (2002) as 209 +/- 4.7 km. Its binarity was recognized in 2002 from adaptive optics (AO) observations (Merline et al., 2002). Soon afterwards, Marchis et al. (2005) conjectured a bilobed "snowman" shape from high resolution imaging with the Keck 10m telescope in the framework of an astrometric AO campaign carried out in 2003 and 2004. Apart from these AO observations, this minor planet, member of the Cybele group, is poorly characterized: it has been observed during only four oppositions between 1976 and 1989 (Debehogne et al., 1978, di Martino et al., 1987, Gil Hutton, 1990, Piironen et al., 1994, De Angelis, 1995, Blanco et al., 1996). Due to the data paucity, it has not been possible to robustly determine the sense of rotation, the sidereal period nor the pole position. The most recent measurement of the synodic period is from 2004 and appears on the website of R. Behrend[1] where the rotation rate was determined as 5.551h with a lightcurve amplitude of 0.45 mag.

From the full determination of the orbital characteristics of Hermione's satellite (Marchis et al., 2005), we derived the orbit pole and adopted it as a provisional pole solution for Hermione ($\lambda_0=1.5\pm2.0°$, $\beta_0=10\pm2.0°$ in ecliptic J2000), based on the absence of any observed or detectable precessional motion of the orbital pole (meaning that we assumed that the satellite orbits in the equatorial plane of the primary). Considering this pole solution, we pointed out that no observations close to the equatorial aspect of the primary had been made so far. Consequently the maximum amplitude should be greater than the measured 0.45 mag in 2004 and the actual shape and size parameters should be revised. On the basis of this pole solution, we predicted that 121 Hermione should reach its equinox in June 2007, fortunately very close to its opposition. We organized a long-term observational campaign for a thorough investigation of its photometric lightcurves. These kinds of photometric measurements are important per se for at least a twofold purpose, i) to determine the asteroid's shape, ii) to refine its rotational properties in terms of period and pole solution. Another reason for observing a binary asteroid system photometrically in an edge-on configuration is the occurrence of mutual events. We successfully estimated the sizes of the main-belt asteroid 22 Kalliope and its moon Linus (Descamps et al., 2008b) via the detection of mutual event in 2007. However, we estimated that in the case of Hermione and S/2001 (121) 1, its moonlet, such events would be barely detectable because their diameter ratio is smaller (~0.08, according the previous measurement made by Marchis et al., 2005).

---
[1] http://obswww.unige.ch/~behrend/page1cou.html#000121

## 2. Observations

**2.1 Photometric observations**

Observations were acquired using small reflectors located in the Southern hemisphere, suited for observing Hermione since in 2007 its declination was -25°. The target magnitude in visible was typically in the range 12-13. At the Makes Observatory (La Réunion Island, observatory code 181), we used a 0.34m (C14) with a CCD SBIG ST-7 camera (765 X 510 pixels) in V filter. All exposures, with typical integration times of 120s, were recorded under atmospheric seeing conditions of 3". Observations at LNA (Laboratório Nacional de Astrofísica, Brazil, observatory code 874) were carried out with the 0.6m Boller & Chivens telescope using an R filter. The CCD used was a Marconi CCD of 2048x2048 pixels, model 42. Pixel scale (not binned) was 0.6 arcsec per pixel. All image processing, including bias subtraction and flat field corrections, was performed using the TASP (*Traitement astrométrique et photométrique*) software package developed by IMCCE. The magnitudes for the asteroid and comparison stars were extracted from each image using the profile fitting photometry technique. Lastly, some observations were made with the University of North Carolina's PROMPT

Gamma Ray Burst optical afterglow observatory at Cerro Panchon in Chile. The PROMPT array currently consists of five 0.41-m Ritchey-Chretien telescopes each outfitted with rapid-readout Alta U47+ cameras by Apogee, which make use of E2V CCDs. Images were obtained with a Johnson R filter and were calibrated and reduced with the image processing program MIRA using aperture photometry. The overall accuracy of relative photometry is better than 0.03 mag.

Aspect data are given in Table 1, including the date of the observation, the geocentric longitude ($\lambda$) and latitude ($\beta$) of the asteroid, its phase angle ($\alpha$) and its geocentric distance in AU (r), the peak-to-peak amplitude of the lightcurve and the telescope aperture. Collected lightcurves are displayed in Fig.1. Lightcurve amplitudes are all greater than 0.55 mag, value recorded near opposition. At larger phase angles, up to 17°, the change in amplitude is significant and reaches an extreme value of 0.7 mag. This change has a twofold origin, the growing of the phase angle and the changing Earth-asteroid aspect. The amplitude increases monotically with the phase angle at a mean rate of 0.007mag/day, typical of C-class asteroids (Zappala et al., 1990).

The Phase Dispersion Minimization (PDM) technique (Stellingwerf, 1978) was used to search for the synodic rotation period within the photometric data. Based on a trial period, PDM bins data according to the rotational phase, where we assumed that two maxima and minima occurred per rotation, the average variance of these subsets is compared to the overall variance of the full set of observations. The best estimate of the period is that for which the ratio of the average variance within a bin to the variance of the sample, which defines the statistic $\theta$, reaches a minimum. This method does not assume any sinusoidal variation of the lightcurve and is well suited for unevenly spaced observations. PDM finds all periodic components or subharmonics (alias periods). Thanks to our six months long span of observations, the periodogram for light-time corrected data of 121 Hermione shows a very salient minimum ($\theta=1.06$) corresponding to the synodic period $P_{syn}$ = 5.55096±0.00015 h (Fig. 2).

**2.2 Adaptive Optics observations**

In addition to AO observations published in Marchis et al., (2005), we collected additional observation with NAOS (Nasmyth AO System) in May 2006 with Yepun-UT4 telescope, one of the 8.2 m telescopes of the Very Large Telescope (European Southern Observatory) located atop of Cerro Paranal. The observations were recorded in direct imaging using the CONICA near-infrared camera equipped with an ALADDIN2 1024x1024 pixel InSb array of 27 μm pixels with the S13 camera (13.27 mas/pixel scale) in broadband filters J, H and K (from 1 to 2.5 μm). NACO, which stands for NAOS-CONICA, provides the best angular correction (60-80 mas) in this wavelength range. More recently, on September 18 2008, an improved version of the Keck II AO system were used to record 2 additional observations with the NIRC2 infrared camera in Jcont narrow band filter (1.21 μm). Due to exceptional seeing conditions and improvement of the AO quality, the angular resolution on these images was 0.035 arcsec.

The basic data processing (sky subtraction, bad-pixel removal, and flat-field correction) applied on all these raw data was performed using the *eclipse* data reduction package (Devillard, 1997). Successive frames taken over a time span of less than 7 min, were combined into one single average image after applying an accurate shift-and-add process through the *Jitter* pipeline offered in the same package.

3. **Shape of Hermione**

**3.1 High amplitude lightcurves of Hermione**

The lightcurve of a spinning body is completely determined by the asteroid shape, light-scattering behavior, and the viewing geometry. Thus the lightcurve of an asteroid can be calculated if one knows its shape, the bidirectional scattering law at every point and the equatorial coordinates of the Sun and Earth. One common simpler shape is a triaxial ellipsoid rotating about its short axis, whose shape is completely parameterized by the ratios b/a and c/a, where a ≥ b ≥ c are the semiaxes lengths. The amplitude of variation at equatorial aspect and zero phase angle is a measure of the ratio of the largest to the intermediate axis as follows:

$$A_{max} = 2.5\log\left(\frac{a}{b}\right) \qquad (1)$$

From the maximum amplitude observed close to opposition, $A_{max}$ = 0.55, we infer an axes ratio a/b=1.6. Quite obviously, such a simplified shape cannot mimic at all the morphology of the collected ligthcurves and ignores much of the conspicuous information that distinguishes individual lightcurves. We know that there is very little albedo variegation on most asteroid surfaces so that we can perform a stringent lightcurve inversion from the bulk of available photometric date in order to assess a more realistic shape model. To perform a thorough investigation of the photometric effect of an irregularly shaped body, it requires numerically modeling the disk-integrated photometric behavior.

**3.2 Modeling the disk-integrated brightness**

The method uses a polyhedron to produce synthetic asteroid lightcurves. Each vertex is given by its (x,y,z) coordinates. In this way, any asteroid shape can be represented by a set of facets. Once the line of sight and the direction of Sun are known, it suffices to select the facets that are both visible by the observer and illuminated by the Sun. The total reflected light is then computed by adding the contribution of each of these *active* facets. In the special case of a concave shape model, self-shadowing occur at non-zero phase angles. This effect, usually not considered in convex bodies, must be taken into consideration because it may represent a non-negligible amount of the overall brightness variation. Each facet reflects the solar light according to its orientation with respect to Earth and Sun but also depending on the adopted scattering law. The two simplest scattering laws used to model planetary surfaces are the Lommel-Seeliger and Lambert law. The Lommel-Seeliger law is a single-scattering law, suitable for low-albedo surfaces. It requires no parameters; it depends solely on the cosines of the incidence and emission angles (the angles between the surface normal and the directions to the light source and to the observer). The Lambert law is a simple description of a perfectly diffuse surface. It is a multiple-scattering law that adequately describes high-albedo surfaces. The Lambert law is $S_L = \mu \mu_0$ and the Lommel-Seeliger law is $S_{LS}=S_L/(\mu+ \mu_0)$, $\mu$ and $\mu_0$ are respectively the cosines of angles of emission and incidence. The adopted bidirectional reflectance law is a linear combination of the Lommel-Seeliger and Lambert scattering models, $S(\mu, \mu_0) = S_{LS}(\mu, \mu_0) + c.S_L(\mu, \mu_0)$, with $c=0.1$.

**3.3 Non-convex shape model of Hermione**

AO imaging made in 2003 with the 10-m Keck telescope early suggested a bilobed shape (Marchis et al., 2005). On 15$^{th}$ January 2005 and 19$^{th}$ September 2008, we performed on the Keck telescope new high resolution observations of 121 Hermione as it was respectively at a geocentric distance of 3.20 and 2.02 AU. Hermione was not far from a pole-on aspect in September 2008. Figure 4 gathers views of Hermione imaged in the near infrared with a resolution up to 35 mas. For most of the images, a "bifurcated" shape with the presence of concavities seems to convincingly favor a non-convex model over a necessarily more artificial convex one. This is why a non-convex inversion was carried out from our new set of photometric observations after having included further lightcurves collected in the past, between 1976 and 2004. In total, 40 lightcurves have been used for the inversion; 2 from Debehogne et al. (1978), 1 from di Martino et al. (1987), 2 from Gil Hutton (1990), 2 from Piironen et al. (1994), 13 collected by Behrend[2] in 2003-2005 and not yet published (circumstances listed in Table 1), and 19 from the present run.

---

[2] http://obswww.unige.ch/~behrend

The general nonconvex shape and spin model was constructed by combining two data modes, photometry and limb/shadow contours from AO images, with the principle described in Kaasalainen and Lamberg (2006) and Kaasalainen (2009). In the AO+LC algorithm, the joint chi-square is minimized with the condition that the separate chi-squares for the two modes be acceptable and mutually optimal. The lightcurve fitting procedure is described in Kaasalainen et al. (2001), and the contour fitting method and the optimal weighting of different data modes are described in detail in Kaasalainen (2009). Nonconvex shape modelling also requires a smoothness constraint (regularizing function) to suppress artificial details in the model, i.e., we chose the simplest model able to fit the data successfully.

As discussed in Kaasalainen (2009), profile and shadow contours (when several viewing angles are available) contain, in fact, almost as much information on the shape and spin as direct images. In adaptive optics, the contours are also more reliable than the actual brightness distributions in deconvolved images, so the modelling is indeed best done by combining contours rather than images with lightcurves. What is more, no scattering model for the surface is needed in the contour mode of AO-data inversion, and the LC-part is not sensitive to the scattering model (Kaasalainen et al. 2001). The procedure is also directly applicable to combining photometry and occultation measurements.

The standard version of the AO+LC code used here employs a function series in spherical harmonics to represent the radii lengths in fixed directions (Kaasalainen and Torppa 2001). In addition to reducing the number of free parameters and providing global continuity, the function series, once determined, gives a representation that can be directly evaluated for any number of radii (or any tessellation scheme) without having to carry out the inversion again. The number of function coefficients, rather than the tessellation density, determines the level of resolution. Strongly bifurcated shapes are not "starlike". A starlike figure can have concavities, but not so extreme that the surface folds back on itself as seen from the center of mass. That is, the radius vector from the center to any point on the surface is single-valued. Convex, or "convex hull", is a figure that has no concavities. The easiest way to think of this is the shape one would have if you put "shrink wrap" around the actual figure. A "non-starlike" figure has concavities so extreme that a radius vector in certain directions is multi-valued. Hermione appears to be a borderline case: the radius representation reproduces the observed AO images well, considering the uncertainties in contour determination. However, the shape model and AO images obviously suggest a somewhat bifurcated body, so the model should be taken as a conservative estimate. The choice of shape representation introduces a bias in the estimation of the degree of bifurcation, so, e.g., a two-lobed contact binary model would always encourage more bifurcated solutions. Contour determination is defined for general shapes in Kaasalainen (2009), and we plan to have a more detailed look into the AO+LC shape solutions (and their biases and systematic errors) for shapes more complex than starlike ones in a future paper.

Figure 1 shows the synthetic lightcurves yielded by the non-convex "bean" shape solution displayed in Fig. 3. They match all available observations. The sidereal period was simultaneously determined in the course of the inversion process, using the synodic period as initial guess, and fitted to $P_{sid}$ = 5.550878±0.000003 h together with the pole direction expressed in J2000 ecliptic coordinates, $\lambda$=3.8 ± 3° and $\beta$=+12.5 ± 10°.

**3.4 Size of Hermione from direct comparison with Adaptive Optics images**

The tridimensional shape model depicted in the previous section was used to pin down the equivalent diameter of Hermione, taken as the equivalent sphere that contains the same volume as the shape model, by scaling it to the contours provided by our high resolution images. Usually, raw images are still slightly fuzzy so that we cannot readily distinguish the silhouette of the object. Therefore, we have applied a laplacian filter in order to enhance the edge by highlighting regions of rapid intensity change. These images were thereafter processed through a morphological analysis to precisely extract the enhanced edges. The morphological treatment applies a gradient operation through a structuring 2-dimensional element or morphological mask which is passed over the image. The final contour is then extracted from the contoured image by taking isophots in the range 2-10% of the maximum brightness in order to precisely catch the object edges highlighted by the morphological processing. The best selected images of Hermione, recorded with the 10-m Keck telescope, are displayed on Fig. 4. For each of them, the aspect of Hermione was computed from our new solution and projected onto the plane of sky. Three contours provided by the shape solution of Hermione and corresponding to three values of the equivalent diameter, 160, 180 and 209 km, have been overlaid on the AO views. It is straightforward to note the striking agreement and accordingly the reliability of our shape solution with the AO images although some features on the model are not confirmed by AO and can be just artifacts of the inversion.

Figure 5 shows the contour profiles corresponding to the best equivalent diameter minimizing the dispersion. The accuracy on the contour determination is about within 0.2 pixel or 2 mas (pixel size = 9.942 mas/pixel) which gives a measurement accuracy of the equivalent diameter of 5 km in 2008, 8 km in 2003 and 10 km in 2005. We adopt as the equivalent diameter of Hermione $D_{eq} = 188 \pm 6$ km, from the best AO images done in September 2008 (resolution of 35 mas, almost twice better than in 2003 and 2005). Other determinations are in agreement with this value within their own uncertainties. This diameter is ~10% smaller than its IMPS diameter of $209 \pm 5$ km, based on radiometric measurements made by the IRAS telescope (Tedesco et al., 2002). Such a similar trend with respect to IMPS data has been recently noted for another main belt binary asteroid, 22 Kalliope (Descamps et al., 2008b).

## 3.5 Estimate of the geometric albedo

Four of the sets of observations in the literature are absolute photometry and can be used to establish a proper albedo for Hermione, using the shape model from lightcurve inversion and scale from AO observations. The maximum V-band brightness at the observed phase angle α, $V_o(\alpha)$ has been estimated. Table 2 gives the tabulated phase angle, $V_o(\alpha)$ and $H_o(G = 0.09)$. For each one the viewing aspect has been reconstructed for the orientation where the synthetic lightcurve is maximum unless otherwise stated. These absolute measures are corrected for the ratio $R_{eff}/R_{eq}$, where $R_{eff}$ is the cross-sectional area:

$$H_{corr} = H_0 + 5\log\left(\frac{R_{eff}}{R_{eq}}\right) \quad (2)$$

where we can take the $R_{eq}$ of 94 km. Finally, adopting G = 0.09, we get a mean $H_o$ value of 7.51 ± 0.2. Since a sphere is the shape with the smallest surface area, the "equivalent sphere" cross-section is very much less than the "average" projected area of the real body, and corresponds to fainter than "average" brightness. This explains why we get a number considerably fainter than the hitherto accepted H value of 121 Hermione (7.31 from Tholen et al., 2007), which is intended to be some sort of "average" brightness at "average" aspect. The geometric albedo we derive, corresponding to an equivalent radius of 94 km and an H magnitude of 7.51, is $p_v$ =0.0495 ± 0.01[3].

## 3.6 Agreement from thermal observations

---

[3] If the size D of an asteroid and its absolute magnitude H are known, the geometric visible albedo $p_v$ can be derived from

$$D(km) = \frac{1329}{\sqrt{p_v}} \cdot 10^{-H/5}$$

To refine the sizes and thus the density estimates of various binary asteroids, we obtained observing time with the InfraRed Spectrometer (IRS) on the Spitzer Space telescope during Cycle-4 (PI: F. Marchis, GO #40164). 121 Hermione was one of the targets on our list and was observed spectroscopically using the low-resolution (R=64-128 from 5.2 to 14.1 μm) and high-resolution modes (R=600, from 9.9 to 19.6 μm) on September 30 2007 at 23:56 UT. The processed thermal spectra were analyzed using the NEATM model (Harris 1998) and are shown in Fig. 6. With H = 7.51, we derived the radiometric effective diameter, the diameter of a sphere having an equivalent projected area, $D_{eff}$ = 164.8 (+8.6 -8.2), $p_v$ = 0.063 (-0.016 +0.020) and a beaming factor $\eta$ = 0.98 (+0.03 -0.01). Using the pole solution, non-convex model and equivalent size ($D_{eq}$ = 188 km) of the primary derived from the AO observations, we can calculate the appearance and projected surface of the primary (see Fig 7) at the time of the Spitzer observations. Including the thermal flux of a 32-km diameter satellite (see section 4 for its determination of its new value), which was within the Spitzer beam, this model predicts an effective diameter of 161.6 km, in agreement with the effective diameter measured with Spitzer/IRS.

Hermione was also observed twice on each of three dates in August – September 1983 by IRAS. We re-analyzed the IRAS data (excluding low-quality data from one sighting) using the NEATM model and derived an effective diameter of D = 204.0 ± 8.8 km. Using our pole solution, the non-convex model and our equivalent size ($D_{eq}$ = 188 km) for the primary, we calculated the projected equivalent diameter for each IRAS sighting. The asteroid was seen close to a pole-on configuration (Hermione sub-Earth latitude of -40°) and our calculated effective diameter (212.1 km) is in very good agreement with the IRAS measurements.

In terms of equivalent diameter, we would have for Spitzer data, $D_{eq}$ =191.7 ± 8.6 km; For IRAS data, 181.8 ± 8.8 km; and for AO, 188.0 ± 6 km. A weighted mean diameter from these three values yields the final adopted equivalent diameter, $D_{eq} = 187 \pm 6$ km.

### 4. Improving the orbit solution of satellite

New AO observations have been carried out in May 2006 on the 8-m VLT-UT4 telescope at ESO-Paranal as well as in September 2008 on the 10-m Keck telescope. The derived astrometric relative positions of the satellite with respect to the center-of-light of Hermione are listed in Table 3. The methodology for measuring positions and fitting a new keplerian orbit solution is described in Marchis et al. (2005). The updated orbital elements are listed in Table 4. The satellite orbit, which is circular, most likely has been circularized by tidal interactions between the satellite and the primary, as seen for various binary asteroids in the main-belt (Marchis et al., 2008). The orbit seems to be equatorial since no precession motion has been detected over 5 years of observations. The root mean square error (RMS) of the whole set of data is of 28 mas in $X$ and 36 mas in $Y$. Furthermore, new measurements of the secondary/primary flux ratio from 2006 and 2008 images clearly indicate that the size ratio was substantially underestimated from earlier observations (Fig. 8). The new inferred size ratio is of 0.17 instead of 0.08, implying a diameter of the satellite of ~32 km, or twice the value derived from past adaptive optics observations (Marchis et al., 2005). A relative low signal-to-noise ratio of adaptive optics images collected in 2003 and 2004 could explain such this discrepancy. The Keck AO system has been improved significantly over the last 4 years and the images are almost twice better resolved on bright target such as 121 Hermione (see Fig. 4 and Fig. 5).

### 5. Revisiting the main physical characteristics

The main refinements are in the semi-major axis and the observed period (Table 3). The total mass of the system $M_T$ comes from the observed period and the knowledge of the $J_2$ which is computed to be 0.28 from the present shape model of Hermione assuming a homogeneous interior. According to the generalized Kepler's third law:

$$n^2 a^3 = GM_T\left(1 + \frac{3}{2}J_2(D_{eq}/2a)^2\right) \qquad (3)$$

Where $G$ is the gravitational constant, n = 140.449 °/day is the sidereal mean motion (the observed angular velocity), a = 747± 11 km the radius of the circular orbit and $D_{eq}$=187 ± 6 km the equivalent diameter of Hermione. We infer $M_T$ = 4.7 ± 0.2 x$10^{18}$ Kg and a bulk density ρ = 1.4 +0.5/-0.2 g/cm$^3$. The error bar accounts for the fact that the real volume might be as much as 20% less than the model volume due to concavities. 121 Hermione is classified as a Ch-type asteroid because of its relatively flat visible spectrum and the presence of a 0.7 μm absorption feature. This feature could be due to the presence of oxidized iron embedded in phyllosilicates formed through aqueous alteration processes (Vilas & Gaffey 1989). It is there very likely that low density hydrated CI & CM chondrites are the best meteorite analogs. They have grain densities around 2.1 g/cm$^3$ (Britt and Consolmagno, 2003) which implies a macroscopic porosity of ~33+5/-20%, close to the adopted lower porosity for loose rubble or soils (Britt et al., 2002). However, more dry or Fe-rich carbonaceous meteorites have densities between 2.9 and 3.5 g/cm$^3$ implying substantially higher macroscopic porosities. Further spectroscopic observations could provide some valuable information about its surface material able to constrain the grain density.

6. Conclusion

We have established the slightly bifurcated shape of Hermione. As a consequence, we have been able to interpret and exploit some past adaptive optics observations allowing us to determine a new value of the primary's equivalent diameter ($D_{eq} \sim 187$ km), which is ~10% smaller than the published IRAS radiometric diameter. We have found that this discrepancy is due to a near-polar view by IRAS, so the thermal flux measurements by IRAS are consistent with our estimates of size within a few percent. After refining the orbit solution of the small 32-km diameter satellite, we have inferred new physical characteristics, such as bulk density (1.4 +0.5/-0.2 g cm$^{-3}$) and macroscopic porosity (~33 +5/-20%).

While the resulting macroporosity of 121 Hermione's primary seems to be equivalent to those of other main-belt binary asteroids, we caution that new careful estimations of the sizes of other main-belt binary asteroids should be made before drawing general inferences regarding their macroporosity. To this end, in those cases in which sufficiently accurate knowledge of the shape and the rotational properties of the primary are available, high resolution imaging with adaptive optics or mutual events photometry, combined with thermal observations, should be pursued to provide new size estimates.

**Acknowledgements**

FM's work was supported by the National Science Foundation Science and Technology Center for Adaptive Optics, and managed by the University of California at Santa Cruz under cooperative agreement No. AST-9876783 and by the National Aeronautics and Space Administration issue through the Science Mission Directorate Research and Analysis Programs number NNX07AP70G.. Part of these data was obtained at the W.M. Keck observatory, which is operated as a scientific partnership between the California Institute of Technology, the University of California and the National Aeronautics and Space Administration. The observatory and its AO system were made possible by the generous financial support of the W. M. Keck Foundation. This work is based in part on observations made with the Spitzer Space Telescope, which is operated by the Jet Propulsion Laboratory, California Institute of Technology under a contract with NASA. Support for this work was provided by NASA through an award issued by JPL/Caltech. JD's work was supported by the grant GACR 205/07/P070 of the Czech grant agency and by the Research Program MSM0021620860 of the Ministry of education. We thank our reviewer, Alan W. Harris for his constructive suggestions which tremendously improved this manuscript.

**Table 1:** 2003-2007 observational circumstances and measured lightcurve amplitudes. The geocentric longitude (λ) and latitude (β) of the asteroid, its phase angle (α) and its geocentric distance in UA (r), the peak-to-peak amplitude of the lightcurve and the telescope aperture are indicated for each epoch of observation.

| Date | λ (°) | β (°) | r (AU) | α (°) | Ampl. (mag) | Telescope (m) |
|---|---|---|---|---|---|---|
| 2003 December 16 | 103.4 | 3.8 | 2.5038 | 7.5 | 0.4 | 0.31 (1) |
| 2003 December 18 | 103.7 | 3.8 | 2.4938 | 6.8 | 0.46 | 0.31 (1) |
| 2003 December 23 | 104.5 | 3.9 | 2.4748 | 5.3 | 0.43 | 0.25 (2) |
| 2003 December 24 | 104.7 | 3.9 | 2.4716 | 4.9 | 0.42 | 0.31 (1) |
| 2003 December 25 | 104.7 | 3.9 | 2.4689 | 4.6 | 0.41 | 0.31 (1) |
| 2004 January 7 | 106.6 | 4.2 | 2.4607 | 1.7 | 0.40 | 1.05 (3) |
| 2004 January 30 | 110.2 | 4.5 | 2.5725 | 7.7 | 0.53 | 0.40 (4) |
| 2004 February 7 | 111.4 | 4.7 | 2.6462 | 9.9 | 0.53 | 0.21 (5) |
| 2004 February 8 | 111.6 | 4.7 | 2.6556 | 10.1 | 0.10 | 0.40 (6) |
| 2004 February. 9 | 111.7 | 4.7 | 2.6661 | 10.3 | 0.40 | 0.40 (6) |
| 2004 February 24 | 114.0 | 4.9 | 2.8462 | 13.4 | 0.58 | 0.31 (1) |
| 2005 February 9 | 160.6 | 7.6 | 2.9584 | 6.7 | 0.03 | 0.25 (7) |
| 2005 February 10 | 160.7 | 7.6 | 2.9524 | 6.5 | 0.03 | 0.25 (7) |
| 2007 March 28 | 271.70 | -0.32 | 3.395 | 16.09 | 0.62 | 0.34 (8) |
| 2007 March 30 | 272.21 | -0.41 | 3.331 | 16.04 | 0.62 | 0.34 (8) |
| 2007 April 13 | 273.51 | -0.74 | 3.113 | 15.37 | 0.70 | 0.34 (8) |
| 2007 April 20 | 273.84 | -0.92 | 3.009 | 14.71 | 0.65 | 0.34 (8) |
| 2007 April 28 | 273.94 | -1.14 | 2.898 | 13.68 | 0.64 | 0.41 (9) |

**Table1 -** Continued

| Date | | | | | | |
|---|---|---|---|---|---|---|
| 2007 May 4 | 273.82 | -1.31 | 2.819 | 12.70 | 0.60 | 0.41 (9) |
| 2007 May 5 | 273.78 | -1.34 | 2.807 | 12.52 | 0.60 | 0.34 (8) |
| 2007 May 19 | 272.77 | -1.76 | 2.652 | 9.49 | 0.63 | 0.41 (9) |
| 2007 June 8 | 269.96 | -2.38 | 2.507 | 3.82 | 0.54 | 0.34 (8) |
| 2007 June 10 | 269.61 | -2.44 | 2.498 | 3.19 | 0.59 | 0.41 (9) |
| 2007 July 2 | 265.64 | -3.03 | 2.479 | 4.19 | 0.60 | 0.41 (9) |
| 2007 July 12 | 264.03 | -3.24 | 2.515 | 7.26 | 0.58 | 0.34 (8) |
| 2007 July 23 | 262.63 | -3.42 | 2.583 | 10.29 | 0.66 | 0.41 (9) |
| 2007 August 11 | 261.49 | -3.63 | 2.757 | 14.31 | 0.68 | 0.34 (8) |
| 2007 September 1 | 262.28 | -3.75 | 3.006 | 16.76 | 0.65 | 0.34 (8) |
| 2007 September 13 | 263.64 | -3.79 | 3.161 | 17.29 | 0.65 | 0.60 (10) |
| 2007 September 15 | 263.92 | -3.80 | 3.187 | 17.32 | 0.70 | 0.60 (10) |
| 2007 September 16 | 264.07 | -3.80 | 3.200 | 17.33 | 0.66 | 0.60 (10) |
| 2007 September 22 | 265.04 | -3.81 | 3.279 | 17.31 | 0.65 | 0.60 (8) |

[1] René Roy: d=0.31m, [2] Laurent Brunetto: d=0.25m, [3] Pic du Midi Observatory: d=1.05m, [4] Stefano Sposetti: d=0.4m, [5] Laurent Bernasconi: d=0.21m [6] Federico Manzini: d=0.40m, [7] Laurent Bernasconi: d=0.25m, [8] Makes Observatory: d=0.34-m, [9] PROMPT Cerro-Tololo : d=0.41-m, [10] LNA : d=0.60-m

**Table 2:** Absolute photometry of 121 Hermione taken in the literature. $R_{eff}$ is the "equivalent sphere" cross-section where the synthetic lightcurve is maximum. A *G* value of 0.09, typical of C-type asteroids, is assumed. $H_0$ is an estimate of H at α=0°. $H_{corr}$ is an estimate of H for the area subtended by the "equivalent sphere" of radius $R_{eq}$=94 km. We obtain a mean H value of 7.51 ± 0.2 mag.

| Date | Phase angle : α(°) | $V_0$ (α) | $H_0$ | $R_{eff}$ | $H_{corr}$ | Reference |
|---|---|---|---|---|---|---|
| 1981/05/06.36 | 4.77 | 7.81 | 7.37 | 97.748 | 7.46 | Zellner et al., 1981 |
| 1983/11/12 | 2.50 | 7.43 | 7.15 | 113.89 | 7.58 | Di Martino et al. 1987 |
| 1985/01/16 | 1.90 | 7.48 | 7.25 | 114.610 | 7.69 | Piironen et al. 1994 |
| 1989/10/26.25 | 9.75 | 7.70 | 7.00 | 116.257 | 7.47 | Gil Hutton 1990 |
| 1992/03/26 | 10.00 | 7.73 | 7.02 | 108.206 | 7.34 | Blanco et al.1996 |

**Table 3:** Relative positions of Hermione' satellite. 2006 Observations were recorded using the NAOS-CONICA Adaptive Optics system available at Yepun-VLT (ESO, Chile) while 2008 observations have been carried out on the 10-m Keck telescope..

| UT Date | UT | X (arcsec) | Y (arcsec) |
|---|---|---|---|
| 5/20/2006 | 3: 7:36. | 0,0658 | 0,3265 |
| 5/20/2006 | 3:20:29 | 0,0658 | 0,3145 |
| 5/23/2006 | 3:28:05 | 0,2770 | 0,0460 |
| 5/27/2006 | 0:59:18 | -0,2749 | 0,0011 |
| 5/28/2006 | 2:08:10 | 0,2228 | 0,1693 |
| 09/19/2008 | 07:32:00 | 0,3460 | -0,3660 |
| 09/19/2008 | 10:10:15 | 0,4270 | -0,2850 |

**Table 4:**
The circular orbit solution derived after adding new astrometric observations made in May 2006 to the set of data published in Marchis et al. (2005). Elements are given in mean J2000 equator and equinox. No eccentricity was inferred.

| Orbital parameters | Best-fit value |
| --- | --- |
| Period (days) | 2.5632 ± 0.0021 |
| Semi-major axis (km) | 747 ± 11 |
| Orbit Pole solution in ECJ2000 (degrees) | $\lambda = 358.7 \pm 4°$, $\beta = 12.5 \pm 4°$ |
| Spin Pole solution in ECJ2000 (degrees) | $\lambda = 3.8 \pm 3°$, $\beta = 12.5 \pm 10°$ |
| Inclination (degrees) | 79.1 ± 4.0 ° |
| Ascending node (degrees) | 83.7 ± 3.0° |

**Figure 1:** Collected lightcurves not corrected for light-time. See Table 1 for details on each observation. Synthetic lightcurves reckoned from the inverted non-convex shape model have been overlapped (solid lines). The moonlet photometric contribution has not been accounted for. Discrepancies are of 0.04 mag at most.

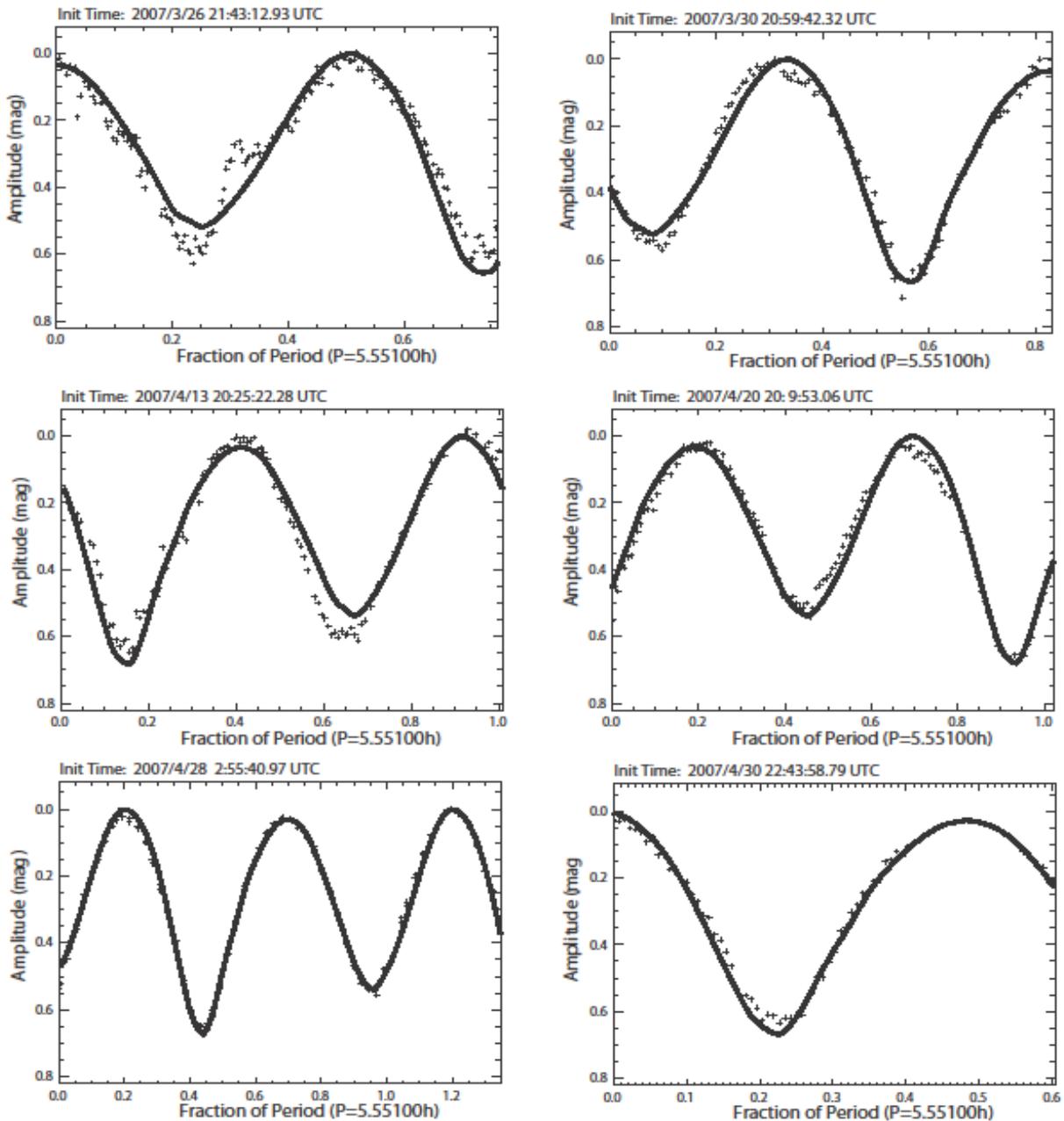

**Figure 1 -** continued

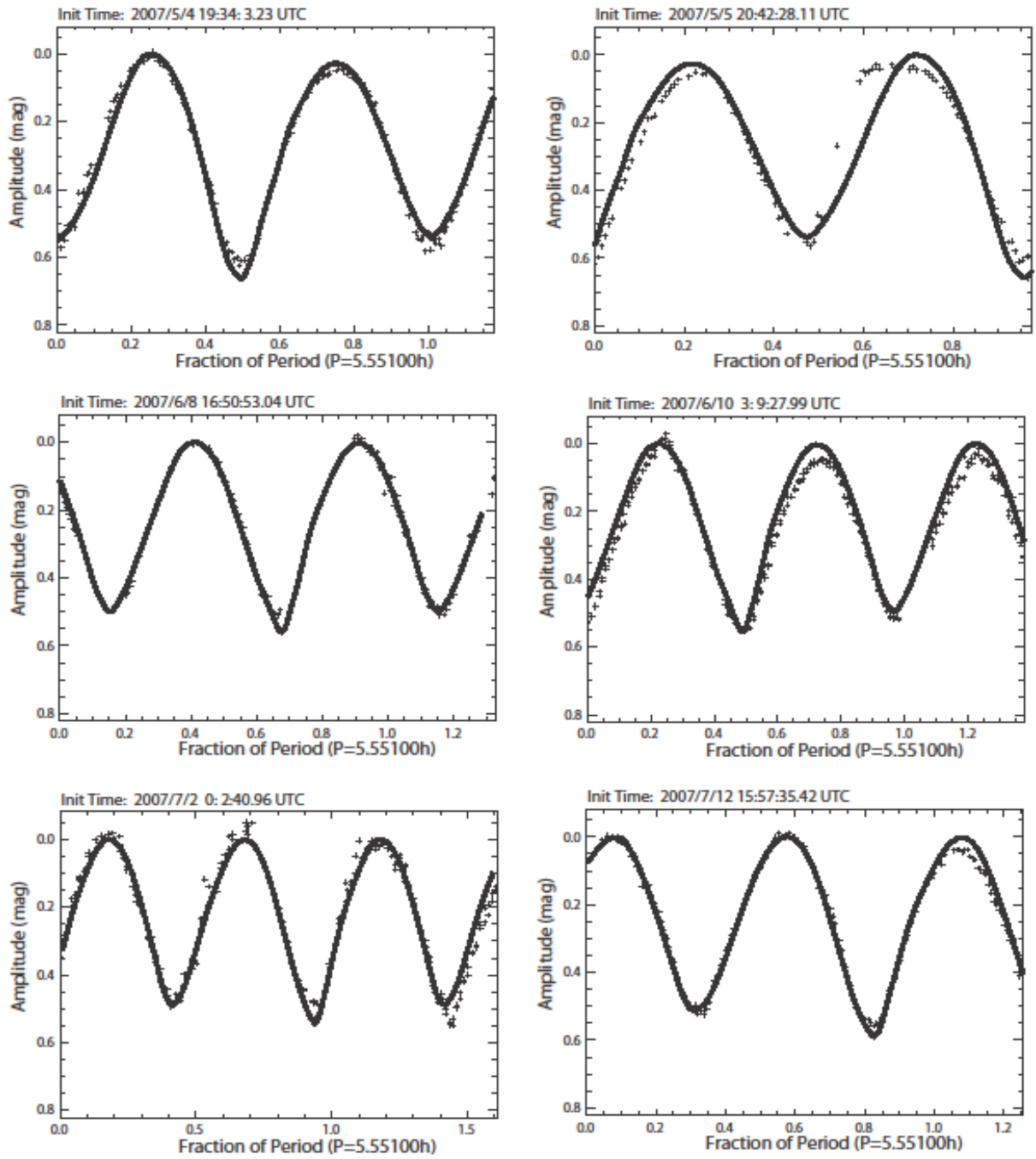

**Figure 1.** Continued

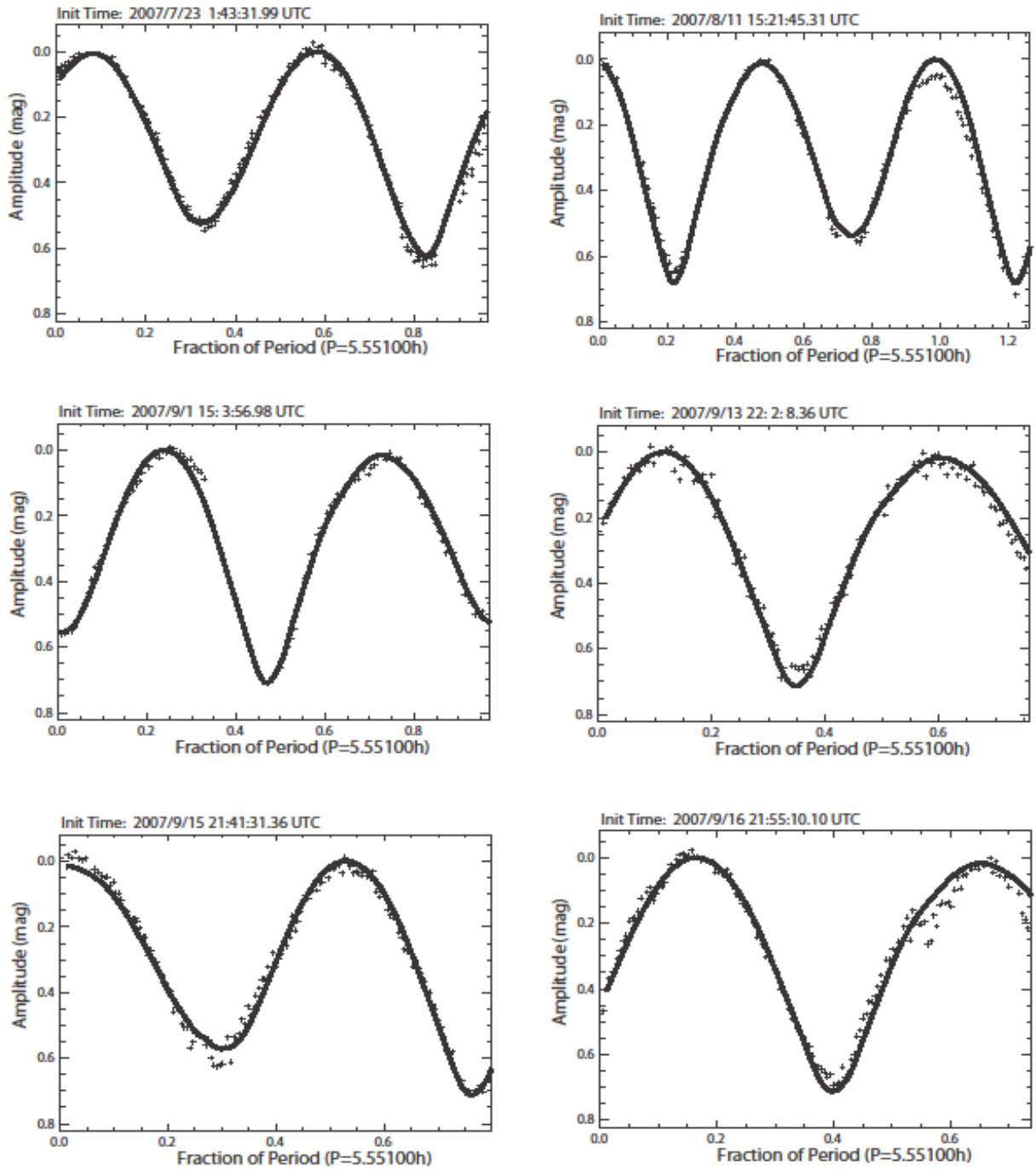

**Figure 1.** Continued. Observations carried out in September 1976 and November 1983 are respectively drawn from Debehogne et al. (1978), Di Martino et al. (1987). Observations of December 2003 and February 2004 have been collected by Raoul Behrend from its website (http://obswww.unige.ch/~behrend).

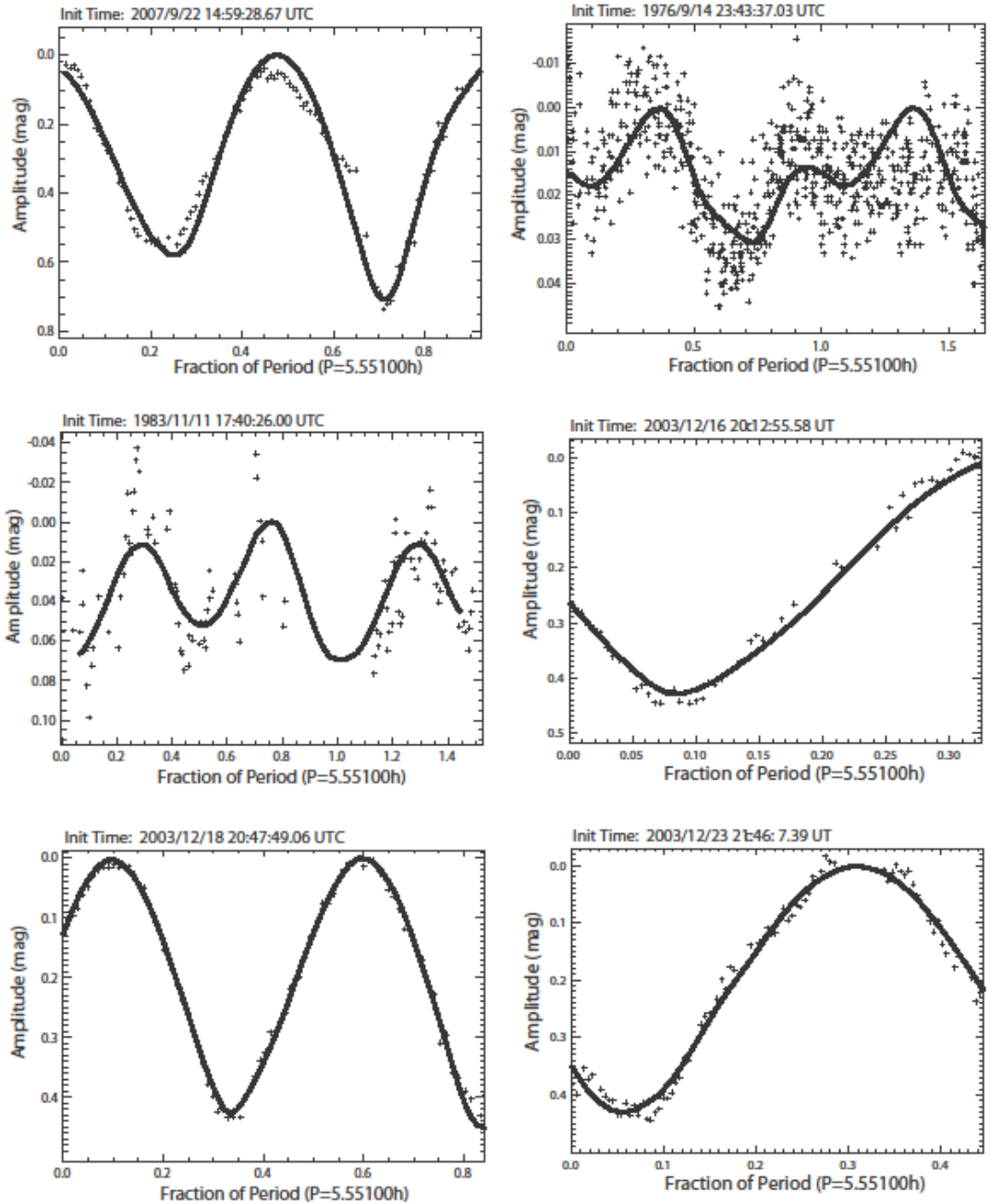

**Figure 1.** Continued.

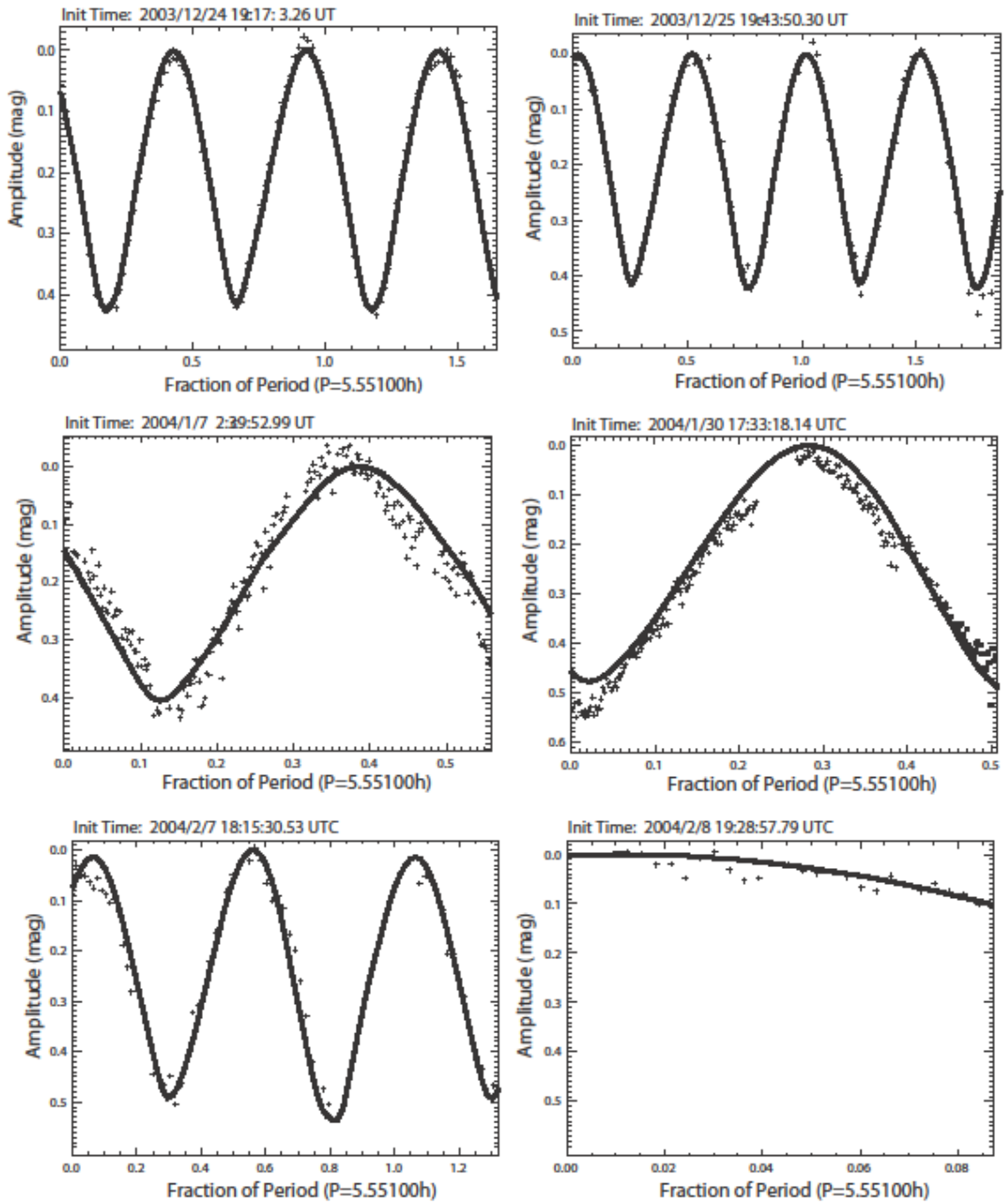

**Figure 1.** Continued.

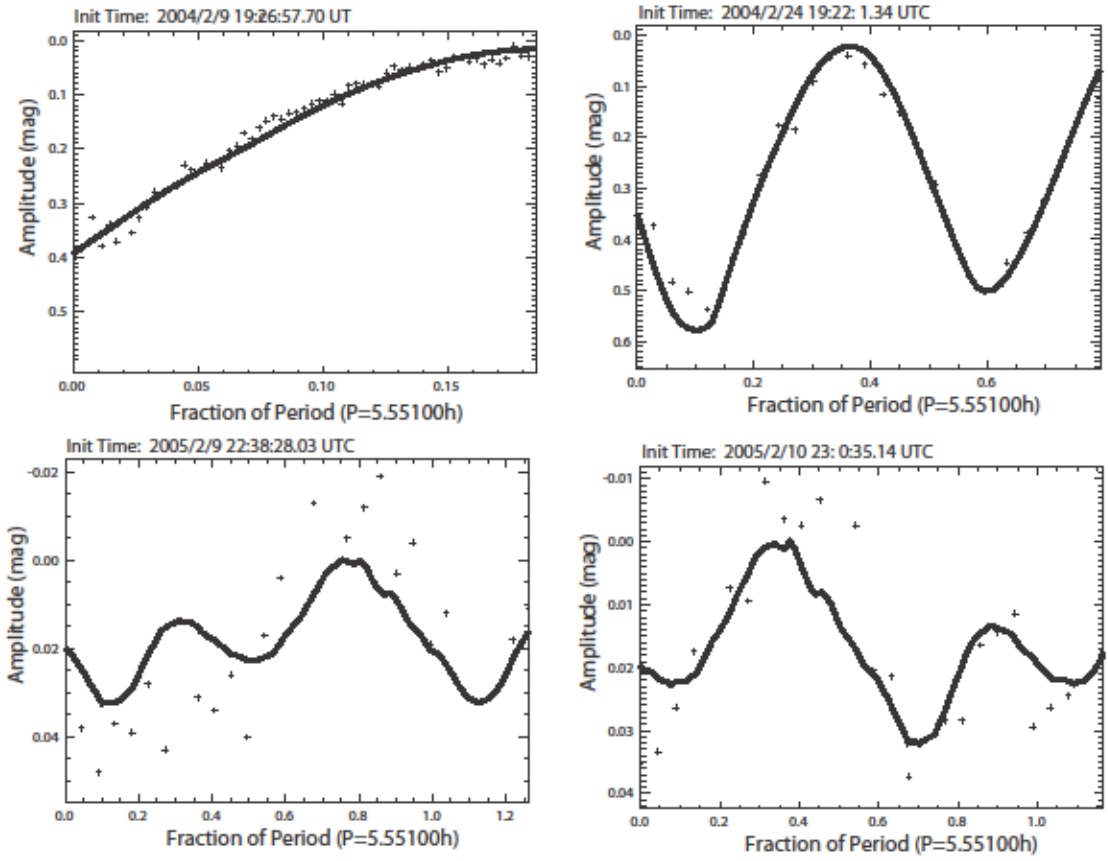

**Figure 2:** PDM plot for the Hermione data showing θ vs. period. The minimum θ value is for P=5.55096 ± 0.00012 h.

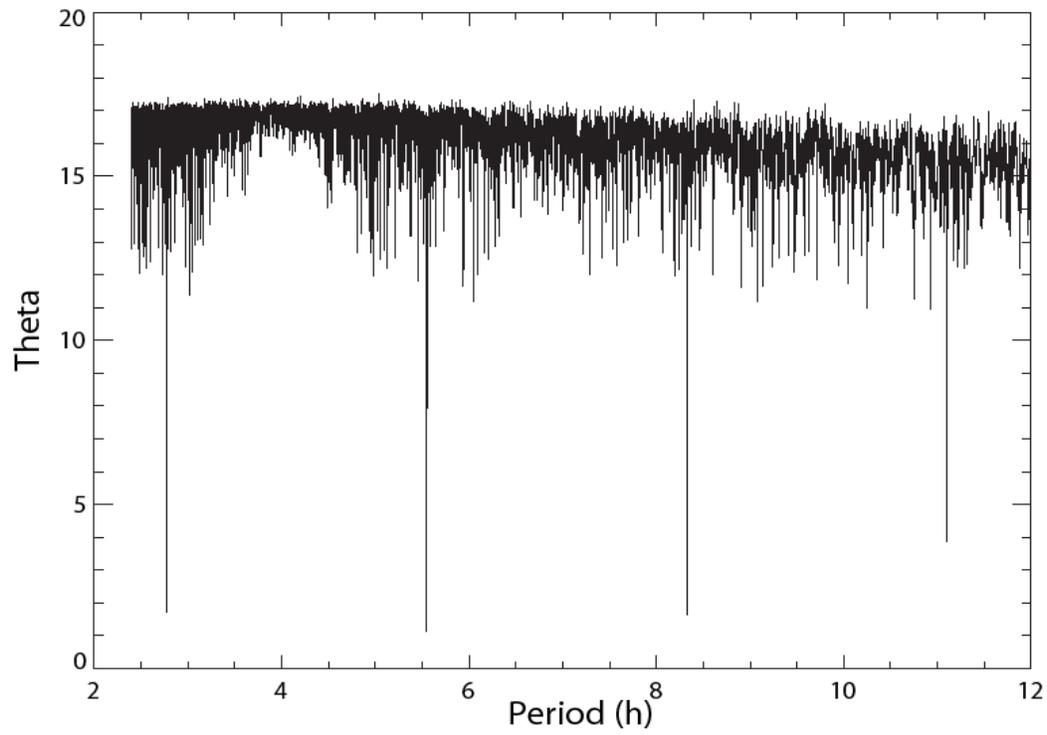

**Figure 3:** Polyhedral non-convex shape solution of 121 Hermione derived using 40 lightcurves observations from 1976 to 2007 and AO images carried out on the 10-m Keck telescope on December 2003, January 2005 and September 2008 (see Fig. 4). The top leftmost picture is a pole-on view of Hermione. Other pictures are edge-on views.

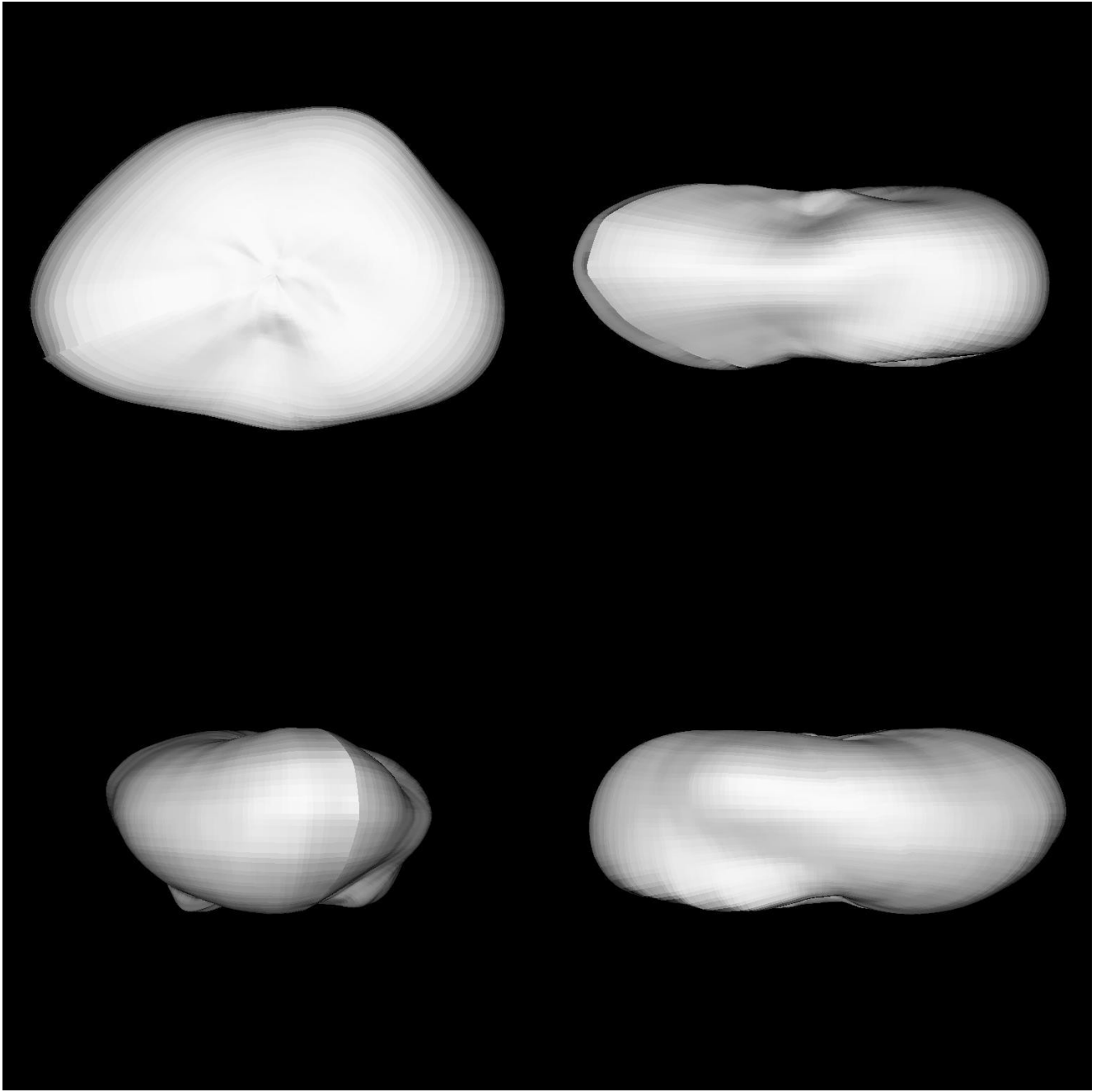

**Figure 4:** Adaptive optics images of Hermione taken with the 10-m Keck telescope in December 2003, January 2005 and September 2008. A laplacian filter has been applied in order to enhance the edge of Hermione. Our non-convex shape solution is shown for the corresponding observing dates. North is up and East is left. The spatial resolution is of about 52 mas, or 95 km in December 2003, 121 km in 2005 and 35 mas or 51 km in 2008. Some features on the model are not confirmed by AO and can be just artifacts of the inversion. Contours of the shape solution corresponding to three equivalent diameters of 160, 180 and 209 km have been superimposed.

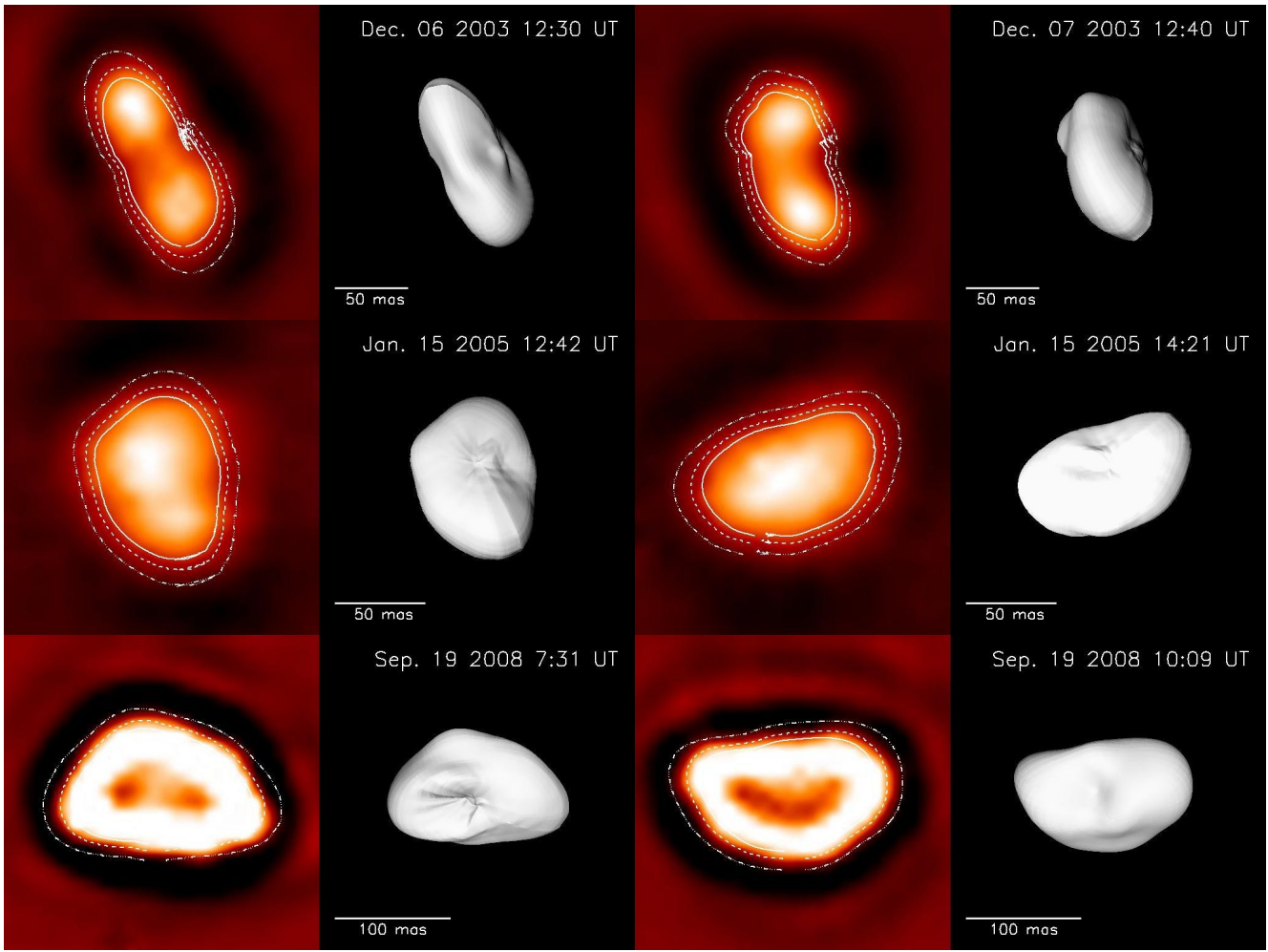

**Figure 5:** Hermione profiles from the AO images of Fig. 4. The contour of resolved Hermione is given in dashed line while the one computed from the projected shape model for the best fitted equivalent diameter appears in solid line. The accuracy on the contour determination is about within 0.2 pixel or 2 mas (pixel size = 9.942 mas/pixel) which gives a measurement accuracy of the equivalent diameter of 5 km in 2008, 8 km in 2003 and 10 km in 2005.

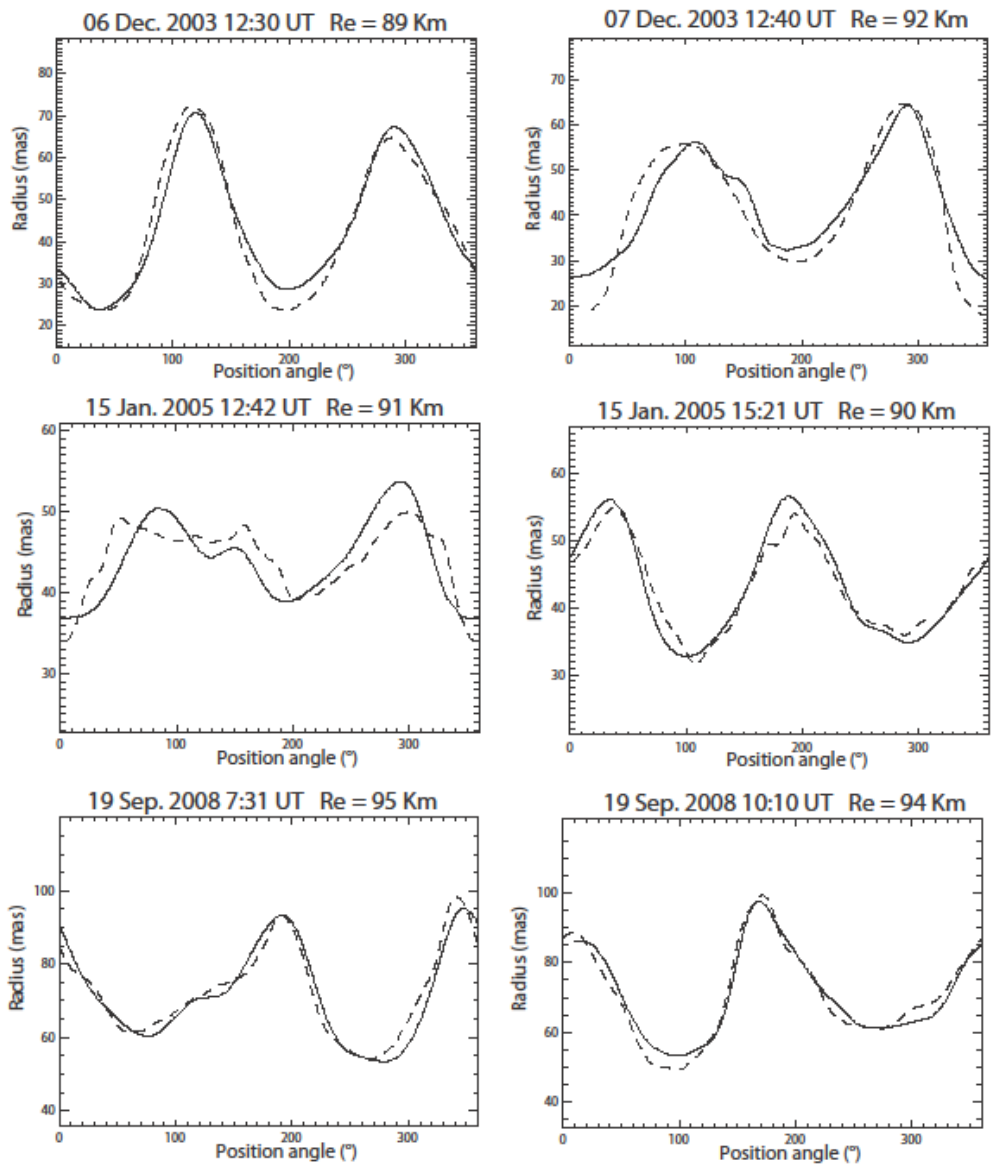

**Figure 6:** Thermal flux low-resolution and high resolution spectra of 121 Hermione recorded using Spitzer/IRS on September 30 2007 at 23:56 UT. We derived the effective diameter, albedo and beaming factor by a fit with the NEATM model.

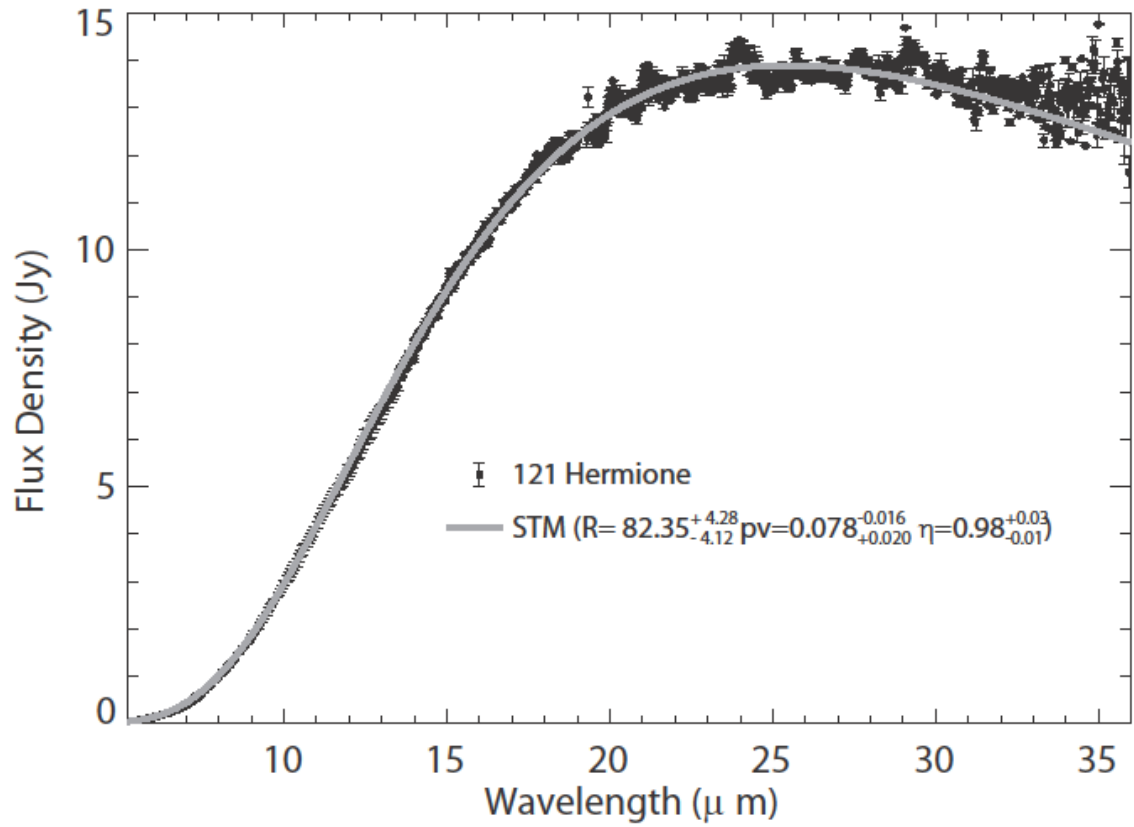

**Figure 7:** Appearance of Hermione on August 18 1983 at 22:25 UT, time of IRAS sighting (left panel) and on September 30 2007 at 23:56 UT, time of Spitzer/IRS observations (right panel). The radiometric effective diameter of Hermione is respectively 211.1 and 161.6 km for an equivalent diameter of 188 km. From thermal measurements, the NEATM model gives an effective diameter of respectively 204.0 ± 8.8 km and 164.8 (+8.6 -8.2) km.

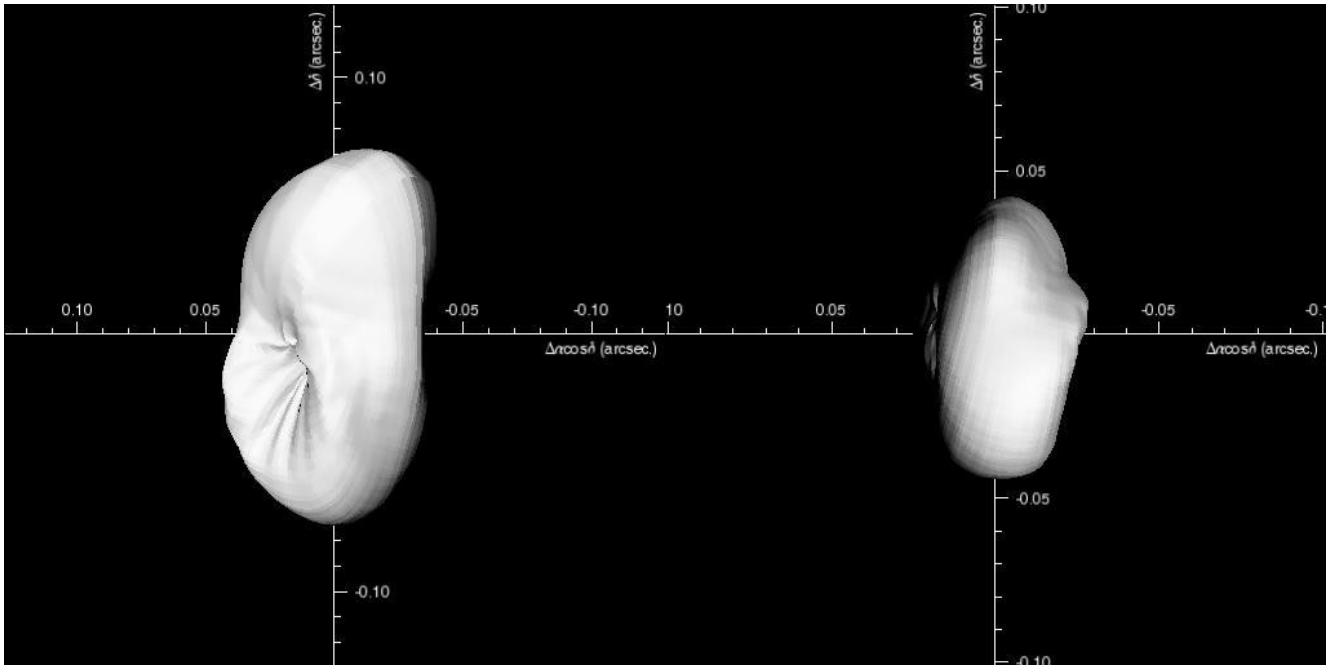

**Figure 8**: The binary system of 121 Hermione observed with the improved Keck II AO system and NIRC2 camera at 1.2 μm on September 19 2008 at 10:10 UTC.

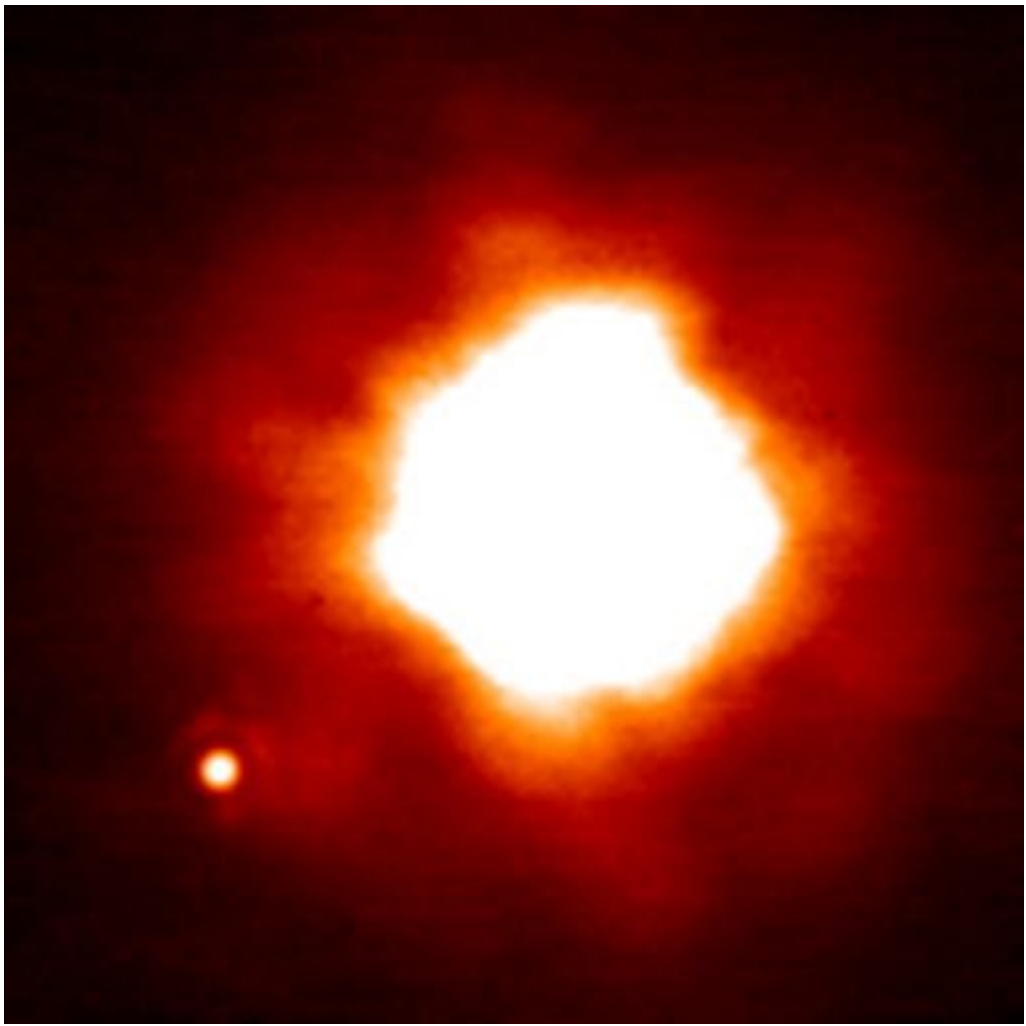